# Charting the Energy Landscape of Metal/Organic Interfaces via Machine Learning

Michael Scherbela, Lukas Hörmann, Andreas Jeindl, Veronika Obersteiner, and Oliver T. Hofmann.

*Institute of Solid State Physics, NAWI Graz, Graz University of Technology,*
*Petersgasse 16, 8010 Graz, Austria*

**The rich polymorphism exhibited by inorganic/organic interfaces is a major challenge for materials design. In this work we present a method to efficiently explore the potential energy surface and predict the formation energies of polymorphs and defects. This is achieved by training a machine learning model on a list of only 100 candidate structures that are evaluated via dispersion-corrected Density Functional Theory (DFT) calculations. We demonstrate the power of this approach for tetracyanoethylene on Ag(100) and explain the anisotropic ordering that is observed experimentally.**

**Introduction.** Without knowing the atomistic structure of not yet synthesized materials, little can be said about their properties. This is a particular problem for organic-based applications, such as organic electronics, where the critical parameters such as electrical conductivity [1] and injection barriers [2] are strongly affected by the interface structure. Before synthesizing a new material, it is therefore highly desirable to computationally screen it for possible polymorphic forms and/or the propensity to form defects that may affect interface properties. However, currently most structure prediction methods are designed for isolated molecules [3] or compact bulk systems. [4,5] Only few approaches deal with interfaces, and also there, with few notable exceptions [6], the target is usually the geometry of isolated adsorbates rather than the polymorphism of extended monolayers. [7–9]

For organic monolayers, often several thousand potential local minima (corresponding to different polymorphs) exist. In practice, the small energy differences between them lead to rich polymorphism and high defect concentrations. [10] Very often, structures with several inequivalent molecules [11,12] are formed. For computational structure prediction, this leads to a fundamental dilemma: While the small energy differences require employing highly-accurate first-principle methods [13], the large unit cells limit their applicability. This is because the large unit cells render each energy evaluation prohibitively expensive, while at the same time, the many degrees of freedom lead to a "combinatorial explosion" of the number of possible structures. Established stochastic methods can therefore only ever explore a tiny fraction of the vast configurational space, potentially missing the ground state structure and giving no systematic overview over possible polymorphs and corresponding defects.

In this contribution, we demonstrate how such an overview can be obtained using a quasi-deterministic, machine-learning based approach. Our approach requires as few as 100 DFT calculations, allowing us to chart the polymorph landscape at affordable cost. We focus the demonstration on the case of tetracyanoethylene (TCNE) adsorbed on Ag(100). This is an ideal "fruit-fly" system, since TCNE is known to form different polymorphs on various metal substrates. [14–16] Moreover, earlier STM experiments indicate that the structure on Ag(100) exhibits a high defect propensity, but only in one crystallographic direction and not the other. In the following, we will first explain our machine learning approach, present a benchmark on a simplified system, and then apply the approach to TCNE/Ag(100). Our overview over the potential energy surface allows us to identify the ground state structure as well as to discuss the similarities and discrepancies between theory and experiment. Furthermore, since our approach yields physically interpretable potential energy maps, we can explain why this unusual, kinked interface structure occurs.

**Predicting the Potential Energy Surface.** We obtain an exhaustive overview over the potential energy surface in three steps: First, we discretize the PES to build a large, exhaustive list of polymorph candidates. Secondly, we define a model that assigns energies to all polymorph candidates. Finally, we train this model using DFT and use it to rank all polymorph candidates.

To create a list of polymorph candidates we use the SAMPLE approach, [17] which is developed for commensurate interfaces where the molecule-substrate interaction dominates over the intermolecular interactions: There, we first determine the geometries that a single, isolated molecule would adopt on the surface using traditional, local geometry optimization starting from different initial positions and orientations. All calculations in this work have been obtained using the FHI-aims [18] code package using the PBE+vdW$^{surf}$ method, where the PBE [19] exchange-correlation functional is augmented with the Tkatchenko-Scheffler [20] method (in its parameterization for surfaces [21]) to account for the missing long-range van-der-Waals interactions. This method has been shown to yield reliable adsorption geometries [22], energies [21], and electronic structures [23]. Further computational details are given in the Supporting Information [24].

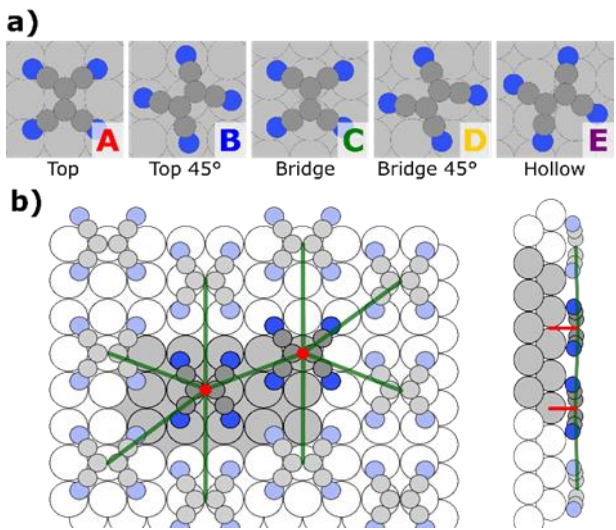

Figure 1: Strategy for structure prediction. a) Local adsorption geometries of TCNE/Ag(100) form the basic building blocks. b) Polymorphs are assembled as combinations of building blocks. Their energies are modelled as interactions with the substrate (red) and pairwise interactions between molecules (green).

For the example of TCNE/Ag(100) we find that the molecule adopts one of five possible adsorption sites, which are depicted in Fig. 1a. We note that four of these structures (A-C and E in Fig. 1a) were previously reported in a different computational study [25], whereas D was not listed there. Conversely, we find two energetically higher-lying geometries reported in ref [25] not to be stable minimum geometries with our methodology.

Secondly, we use these *local adsorption geometries* (and the geometries that are symmetry equivalent by rotation, mirror, inversion and translation) on the substrate as building blocks to assemble larger structures containing multiple molecules/unit cell (UC) (Fig. 1b). This is effectively done by listing all possible combinations of all *local adsorption geometries* on all possible adsorption sites within a given supercell where the molecules do not collide, i.e. are farther apart than a given threshold ($d_{min} = 2.6$ Å). This procedure eliminates unphysical structures and allows a unique, exhaustive enumeration of the many potential energy minima. We note that for our example of TCNE/Ag(100), we find approx. 200.000 possible polymorphs containing up to 8 molecules/UC (see below). Thus, we have only reduced the search space from "completely intractable" to "still too many to be sampled exhaustively".

While this discretization is already useful for finding the ground state structure when combining it with stochastic optimization methods [17], here we want to obtain a more comprehensive overview over the structural space. For this we need an efficient and accurate energy model. Here, it is possible to rely on a simple model, where the formation energy of any structure is given by two sets of energies: Interactions of the molecules with the substrate and interactions between the molecules, as depicted in Fig. 1b. For the molecule-substrate-interaction we introduce one parameter $U_i$ per *local adsorption geometry*. In the specific case of TCNE, there are 5 parameters corresponding to the structures A-E from Figure 1a. For the molecule-molecule-interaction we assign one energy $V_p$ to every possible pairwise interaction between molecules:

$$E_{config} = \sum_{mol\ i} n_i U_i + \sum_{pairs\ p} n_p V_p = n^T \omega \quad (1)$$

The index *p* encodes the interaction between local adsorption geometries (*i,j*) at a given distance *r* (see below). We note that the distances are defined on our discretized grid. Thus, we obtain a different $V_p$ for every different interaction. Also, because the $V_p$ are not (explicit) functions of nuclear coordinates, the different $V_p$ are not analytically connected, and a priori eq. 1 does not hold any information for molecular geometries that are "off" the grid (i.e., where the molecules would be moved to positions that are not local adsorption geometries). If all $U_i$ and $V_p$ were known, the energy of any configuration could be determined by counting the number of occurrences $n_i$ of each local geometry and the number of occurrences $n_p$ of each pairwise interaction. For simplicity, we collect the $U_i$ and $V_p$ in a joint vector representation $\omega$ (see right-hand side of eq. 1).

In principle, one could exhaustively calculate the interactions $V_p$ directly by performing DFT calculations for all pairs of molecules. However, this is impractical for several reasons: Foremost, the number of relevant pairs is very large, requiring immense computational effort. Packwood at al., who used a similar energy model on a discretized grid, suggested to circumvent this problem by calculating only some of the pairwise interactions and use machine learning to predict the rest. [6] However, explicitly calculating specific pairs requires large supercells to decouple each pair from its periodic replicas. For cells of this size, accounting for the substrate becomes intractably expensive. A possible solution to this problem is to omit the substrate and focus only on the interactions between the molecules in their gas phase electronic structure. Despite its success in ref [6], in general this approach bears the risk of missing substrate-induced interactions, especially when (partially) covalent bonds are formed or when charge-transfer occurs between the substrate and the adsorbate. Both is the case for TCNE on Ag(100). Secondly, even if the electronic structure of the adsorbate was correctly accounted for (e.g. by charging the adsorbate), the interactions obtained in this way may differ from the interactions within the system one is ultimately interested in. For instance, in a closed-packed structure, depolarization decreases the electrostatic repulsion between two charged molecules [26].

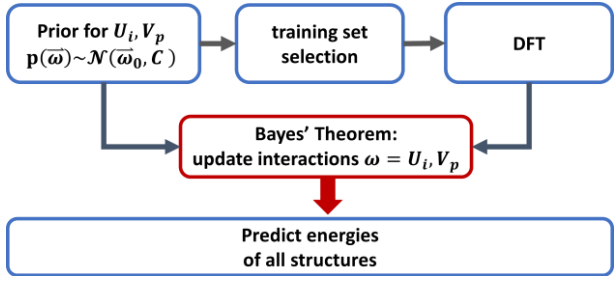

Figure 2: Flowchart of the machine-learning approach

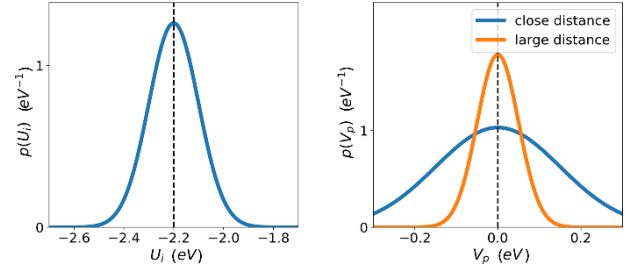

Figure 3: Schematic probability distribution functions for Ui's (left) and Vp's (right) at close and large distance

We circumvent all these issues by not calculating the interactions directly, but rather infer them from selected calculations of the actual, closely-packed structural candidates using Bayes' Theorem. To this aim, we assign a *prior* Gaussian probability distribution to the set of parameters (see below).

$$p(\omega) \propto \exp\left(-\frac{1}{2}(\omega-\omega_0)^T C_0^{-1}(\omega-\omega_0)\right) \quad (2)$$

We then update the probabilities using selected DFT calculations. Finally, we assign each parameter its most likely values based on the *posterior* distribution (see Figure 2).

Initially, we can make the following assumptions about the *prior* probability distribution. For the $U_i$, since we obtained the geometries of the isolated molecules with DFT in the first step of the SAMPLE approach, we already know those individual adsorption energies. These are used as educated guess for the mean of $U_i$. In the closed-packed layer, these may change by a few 10 meV due to depolarization and other effects.

Unfortunately, no comparable information exists about the interactions $V_p$. Our initial guess for the interaction energies between molecules is non-interacting ($V_p = 0$). This guess is likely to be good when the molecules are well separated and less well founded when the molecules are very close. We encode this varying certainty about our initial guess as a different variance $C_{0,ii}$ for each pair of molecules depending on their minimal separation $d$.

$$\sqrt{C_{0,ii}} = \sigma_{pairs}\, e^{-\frac{d}{\lambda}} \quad (3)$$

Here, $\sigma_{pairs}$ is the expected energy range of the interactions (about 100 meV) and $\lambda$ is the length scale at which these interactions decay (exemplarily probability distributions are shown in Figure 3). Furthermore, "similar" pairs of molecules have similar interaction energies, i.e. the interaction potential varies smoothly on our grid. To measure the similarity between pairs the L1 norm of the difference of their feature vectors $v_i, v_j$ was chosen.

$$C_{0,ij} = \sqrt{C_{ii}C_{jj}}\, e^{-\frac{\|v_i-v_j\|_1}{\alpha}} \quad (4)$$

For the feature vectors $v$ we used a sorted list of inverse interatomic distances squared. Only the distances between the "cornerstones" of the molecules (for TCNE, the 4 nitrogen atoms) were chosen, since they already contains all the relevant information.

$$v = \begin{pmatrix} d_1^{-2} \\ d_2^{-2} \\ \vdots \end{pmatrix} \quad (5)$$

This choice of inverse distances leads to a strongly varying potential at small distances and a smooth potential at large separations.

After the prior guess has been constructed, we update the probability distributions of our parameters according to Bayes' Theorem:

$$C\boldsymbol{\omega} = C_0^{-1}\boldsymbol{\omega_0} + N^T \boldsymbol{E_{DFT}}/\gamma^2 \quad (6a)$$

$$C := C_0^{-1} + N^T N/\gamma^2 \quad (6b)$$

Here, $\boldsymbol{\omega_0}$ and $C_0$ are the parameters of the prior distribution (indicated by the index 0). $\boldsymbol{E_{DFT}}$ is a vector of all energies of polymorph candidates that were calculated. $\gamma$ is their accuracy, describing how well the two-body interaction approximation holds. N is a matrix of vectors n that describes how often which parameter occurs in a given polymorph candidate. $C$ and $\boldsymbol{\omega}$ (without indices) are the *posterior* covariance and mean values, i.e. the values assigned by the model after learning has taken place.

We note that this approach contains various free hyperparameters: $\lambda$, $\alpha$, and $\gamma$. However, these have a clear physical interpretation, allowing us to choose sensible values without meticulously optimizing the parameter space. We have chosen the following hyperparameters: $\lambda$ was set to 5Å (slow decay of the distance, to capture long-range effects on the surface) and $\alpha$ was set to 0.3 (medium to weak correlation between the interaction parameters). Our tests indicate that $\gamma$ always small, typically a few meV (We used $\gamma = 5$ meV throughout). In principle, the prediction accuracy might be further improved by systematically optimizing these hyperparameters. However, we found no significant improvements in prediction accuracies when varying these parameters within physically reasonable ranges (i.e., for a given training set size, the RMSE-values remain in the same order of magnitude also when changing the hyperparameters); furthermore, we found these parameters to also work well for two other, conceptually very different systems (naphthalene on Cu(111) and benzoquinone on Ag(111), see below).

The main aspect that governs the efficiency of our machine learning model is a prudent selection of an appropriate training set $E_{DFT}$. In most machine learning applications the training of the model is done after a training dataset has been acquired. This is in particular the case when benchmarking new machine learning models on existing datasets, such as the MNIST database for image classification of handwriting or the QM7 dataset for the atomization energies of small molecules. On the contrary, when searching for low energy structures of a specific system, training data is usually not available and must be supplied by the user. Choice of the training structures could be – and often is - done randomly. This is a prudent choice when the energy function is unknown (e.g., when training neural networks) or when all data are similar (e.g., when specifically learning interactions individually).

In the present case, however, where we train on close-packed structures with multiple different interactions at the same time, it is possible and highly advantageous to select a training set that contains the data-points which offer the "highest gain of information": The goal of Gaussian Process Regression for our application is to accurately estimate the fit coefficients $\omega$, which in turn will allow accurate prediction of energies of all configurations. This is equivalent to minimizing the posterior covariance matrix C. Since C is a matrix, there is no unique definition of minimization. Instead there are a variety of different optimality-criteria defined in the field of Optimal Design Theory. One popular criterion is the so-called d-optimality, which minimizes the determinant of C. Here, we perform this minimization using Fedorov's algorithm [27] which starts with a random subset selection and iteratively improves this initial guess by greedily swapping training samples if this swap decreases the determinant of C. Since the final set of structures depends only on the chosen model and the prior assumptions, it can be selected as a whole before performing any calculations, allowing all DFT calculations to be run in parallel. For a fixed training set size, selecting the training set according to d-optimality notably improves the training accuracy, as shown in Supporting Material [24].

An additional part of our strategy is based on the fact that not all data points are equally costly to acquire. For a given coverage, systems with fewer molecules per unit cell are modelled in smaller unit cells. On a formal basis, DFT scales with the number of electrons in the system cubed [28] and even in practice the scaling is somewhat worse than linear [18,29]. The overall effort can thus be greatly reduced by preferentially sampling systems with a high translational symmetry, which can be modelled in small unit cells, even if the information gain per calculation is smaller. In practice, we realize this by choosing training samples in batches of 10 (and then updating our model according to eq. 6), beginning with small unit cells. Once the prediction accuracy (determined by cross-validation) has fallen below 10 meV, we continue with batches containing larger cells. As we show in the following paragraphs, this selection strategy makes our model particular efficient.

**Benchmarking the Machine Learning Model.** Before tackling the actual system of interest (TCNE on Ag(100)), we need to ask two key questions: 'What prediction accuracy can we obtain?' And: 'Is it indeed possible to predict the energies of large unit cells by training the model only on cheaper, smaller ones?' In principle, this can be done by calculating a reasonably large training set and then using leave-one-out-cross validation (which we also do below). However, we aim for a more comprehensive picture. Thus, to answer these questions, we first benchmark our approach on a well-controlled test system where a more extensive dataset of DFT calculations can be readily obtained.

For this, we consider a hypothetical TCNE monolayer without the substrate, but using the same polymorphs candidates that would also be obtained on the Ag(100) surface. This makes the calculations sufficiently cheap to allow calculating DFT energies on a quasi-comprehensive set of polymorphs candidates. When generating a list of all possible configurations that have the same coverage as observed experimentally on Ag(100) (see Methods Section), we find 251 "small" configurations that contain 2 or 4 molecules/UC, and approximately $2 \times 10^5$ "large" configurations containing 6 or 8 molecules/UC. To reliably assess the performance of our approach, we compiled a reference set that consists of all polymorphs with 6 or fewer molecules/UC, plus 2000 polymorphs drawn randomly from all polymorphs with 8 molecules/UC. The total energies of all ≈6000 of these geometries were calculated using DFT. As discussed in the next paragraphs, we then trained our model on various systematically selected subsets of this dataset to assess its predictions for various training set selections. We emphasize at this point that this hypothetical layer is only used for benchmarking (because it is cheap to calculate), and the results obtained here never enter the calculations for TCNE on Ag(100) discussed in the next section.

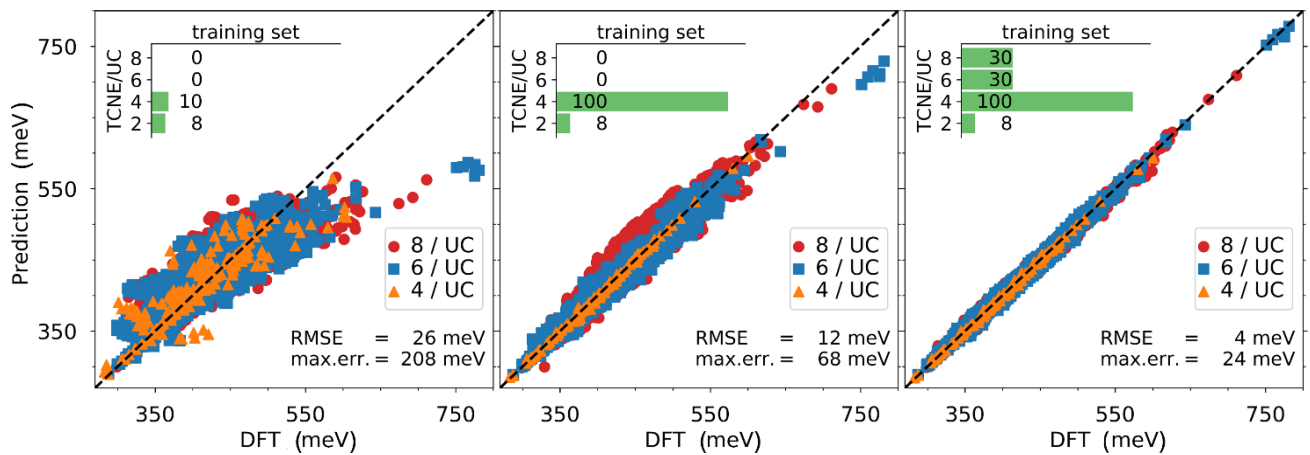

*Figure 4: Comparison between the monolayer formation energies (energy of the total system minus the energies of the isolated TCNE molecules) predicted by the machine learning model and the DFT reference calculations for varying training set sizes for the model system of a free-standing TCNE monolayer. Test points are colored according to the number of molecules within the unit cell. Inset: Number of training samples chosen from each unit cell size. a) shows the situation for a training set of 18 small systems, b) for 108 small systems, and c) for 168 systems, mixed between smaller and larger unit cells..*

Fig. 4 shows the predicted versus the DFT calculated formation energies for the comprehensive data set encompassing all ≈6000 configurations. The panels a-c show the performance of the machine learning model for different training set sizes. In Fig. 4a, the model has seen very few training data, i.e. only 8 polymorph candidates with 2 molecules per cell and 10 polymorphs with 4 molecules per cell. It is therefore still biased towards the initial, non-interacting *prior* guess. Training on these 18 DFT calculations yields a Root Mean Square Error (RMSE) of 26 meV/molecule. Fig. 4b shows the prediction when including only a few more calculations on configurations with 4 molecules per cell (108 in total). It is particularly noteworthy that even though the model has been trained only on some of the small configurations, it gives not only excellent prediction accuracy for similar, small configurations (orange triangles, RMSE = 2.6 meV/molecule), but also yields good accuracy for the datasets with large configurations which it has never been trained on (blue squares and red circles, RMSE = 12 meV/molecule). Since we have performed exhaustive DFT calculations for this model system, we can also confirm that there are no significant outliers (maximum deviation 68 meV). Additionally including a few large configurations into the training set (Fig. 4c) yields a RMSE of 4 meV/molecule across the entire dataset. We emphasize that these energy uncertainties are significantly lower than $k_B T$ at 300K (= 25 meV), or the often quoted "chemical accuracy" of 1 kcal/mol (43 meV) and are even within the numerical accuracy of our DFT calculations, which is approximately 10 meV (see Method Section in the Supporting Information [24]). Such small residual errors are often associated with overfitting, implying that the data is too strongly trained to the test set and unable to predict new, previously unknown data. We would thus like to emphasize that these RMSE values were obtained on a comprehensive test set of approximately 6000 structures and that none of the structures in the test set were at any time part of the training set. We attribute this good performance to the inclusion of prior knowledge and conscious selection of the training set using d-optimality.

Our model is thus indeed able to predict energies with the same accuracy as DFT after having been trained only on 100-200 calculations, which is much more efficient than comparable approaches. [30–33]. Since these calculations preferentially include small unit cells that are computationally cheap, while still allowing predictions of larger, significantly more expensive calculations, the computational effort is reduced by 3-4 orders of magnitude compared to exhaustively calculating all polymorphs.

An obvious question at this point is, however, whether similarly good results could also be expected for other systems. Although we cannot provide a full, comprehensive set of tests, in Figure 5 we provide an overview over the performance for conceptionally very different molecules: naphthalene and benzoquinone. The former is an inert aromatic hydrocarbon, while the latter is a strong, quinoidal electron acceptor with positively charged hydrogens and negatively charged oxygen atoms at its rim. We emphasize that for the training of the model, the same type of hyper parameters and feature vectors have been chosen. Nonetheless, we find the prediction error drops similarly quickly with the size of the training set, indicating that our model is (at least reasonably) transferrable to other organic molecules.

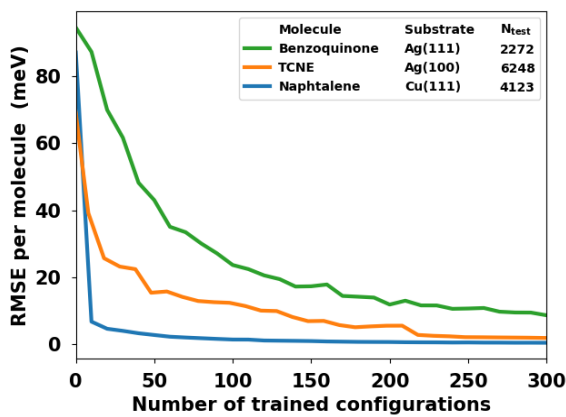

*Figure 5: Evolution of the root-mean-square-error (RMSE) with system size for hypothetical free-standing monolayer, derived for the adsorption of TCNE on Ag(100) as well as two complimentary systems: naphtalene on Cu(111) and benzoquinone on Ag(111). Learning was performed with d-optimal selected sets from a pre-computed test set containing $N_{Test}$ configurations. For TCNE on Ag(100), the pre-computed set exhaustively contains all polymorph candidates with up to 6 molecules per unit cell. The RMSE was computed on the remaining configurations, excluding the training set. For all systems, the same hyperparameters were employed.*

**Application to TCNE/Ag(100)**. With confidence in the performance of our machine-learning approach, we can now turn to the actual system of interest, the closed-packed interface of TCNE on Ag(100).

To predict the potential energy landscape of TCNE on Ag(100) the same training strategy as above was employed, i.e. we select a quasi-deterministic training set of 108 polymorphs (8 polymorphs with 2 TCNE/UC and 100 polymorphs with 4 TCNE/UC), according to d-optimality. We emphasize that this training is completely independent from the above benchmark, i.e. no results from the hypothetical, free standing molecules enter the training of the machine learning model. (Since the electronic structure of TCNE on the surface is fundamentally different than on the surface, these might distort the results.) After training the model on this small dataset the formation energies for all other $2 \times 10^5$ configurations were predicted, allowing a ranking of the configurations according to their predicted formation energies as depicted in Fig. 6. Calculating all these formation energies with DFT would have consumed about 1 million CPU-years on a BlueGene/Q cluster, while calculating the training set required only 0.002% of that effort.

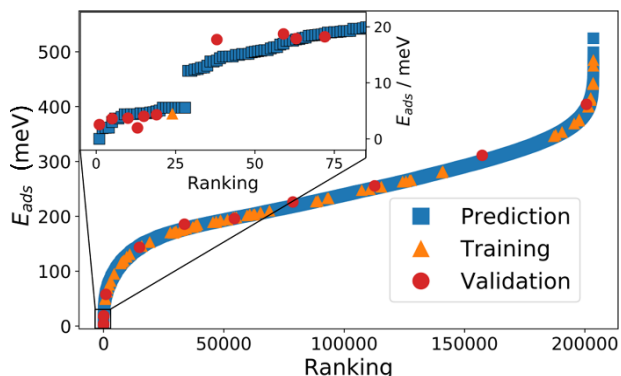

*Figure 6 : Ranking of configurations by predicted formation energies. The inset shows a zoom into the lowest 25 meV. More than 100 configurations lie within this energy range.*

Here, of course, computing an exhaustive dataset to evaluate the performance of the machine-learning model compared to the DFT data is prohibitively expensive. Instead, we additionally selected a sample of 10 polymorph candidates randomly and independently of the training set. Comparing the DFT results with the machine-learning prediction shows a low RMSE of 6 meV/molecule across the entire energy range. Additionally, we specifically validated the prediction for the low-energy region by selecting 8 configurations (that we not in the training set), and find that here, the RMSE even lies at 2 meV/molecule. The accuracy is thus again well within the numerical accuracy of the underlying DFT calculations of approximately 10 meV.

Having finally obtained a comprehensive list of energies for polymorph candidates at DFT accuracy, we can now analyze the structural properties of TCNE on Ag(100).

**Structural Properties.** When designing materials, two important questions are (i) what the ground state structure is, and (ii) whether the material is prone to polymorphism and/or defect formation. For the latter, TCNE/Ag(100) is a particular interesting test system, since earlier STM experiment indicate that it forms a structure with high translational symmetry in one direction, but kinks and periodicities of varying length in the other [16] (Fig. 7a). This is particularly surprising, because the Ag(100) unit cell has a $C_4$ symmetry, i.e., both direction are equivalent. Since also the TCNE molecule has approximately equal dimensions in x and y, one might expect a more isotropic structure, as is also found for TCNE/Cu(100). [14]

Our machine learning algorithms predicts that the structure lowest in energy contains 6 TCNE/UC and consists of diagonal lines of molecules alternating between *top* and *bridge* positions (Figure 7b). We note that only 2 of these molecules are inequivalent, as the same structure can also be described as a monoclinic unit cell containing 2 molecules. The fact that we could correctly find and predict the corresponding rectangular supercell, which is three times as large, is a further sign of the capability of our approach to deal with the vast configurational space.

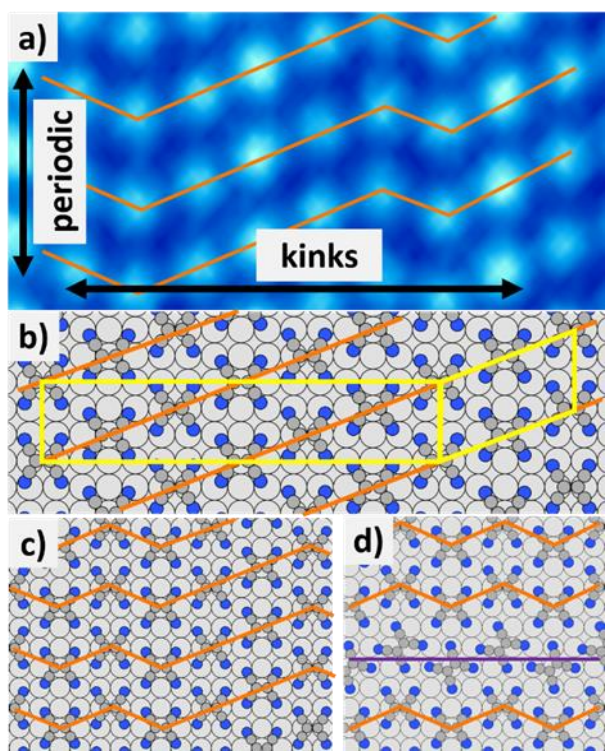

*Figure 7: Exemplary structures predicted by the machine learning approach a) STM image of TCNE/Ag(100) courtesy of Daniel Wegner: Molecules form diagonal lines with strong periodicity in one direction, but frequent kinks in the other direction. b) Computed groundstate: Diagonal lines of molecules, alternating between top and bridge sites. Monoclinic unit cell and rectangular unit cell are marked in yellow. c) Kinks in the diagonal lines are low energy defects (4 meV/TCNE). d) Breaking the periodicity perpendicular to the lines is a high energy defect, requiring 120 meV/TCNE in the defect line.*

Furthermore, we find that there are about 100 configurations within 25 meV/molecule of the predicted global minimum. These are low-energy defect structures that lead to "kinks" in the diagonal structure along one crystallographic direction but not the other. The low energy differences to the ground state indicates that a large variety of different structures are present at room temperature (where the sample has been prepared), which leads to the observed, irregularly kinked layers.

When comparing our predicted ground-state structure to the experimental interpretation drawn from STM and STS measurements [16], we also find notable differences: Experiments report up to four inequivalent molecules per unit cell: Of these, one molecule is clearly assigned to a "top" adsorption site and two are clearly positioned at a "bridge" site. The fourth molecule is more ambiguous, but tentatively also assigned a bridge position in ref [16]. This is potentially at variance with our DFT results, that find a *top*/*bridge* ratio of 1:1 to be energetically more favorable by 250 meV / molecule. We note that this discrepancy is clearly not a consequence of neglecting the vibrational contributions: the difference in between the zero-point energy for "top" and "bridge" amounts only to approximately 20 meV. One may also be inclined to think that the difference may stem from the choice of the functional. We have re-evaluated the local adsorption energies with different functionals (including revPBE, SCAN, and HSE06, see Table I in the Supporting Information [24]), and indeed found differences on the order of 100 meV. This, however, is still too small to change the composition, i.e. the 1:1 top/bridge ratio is a stable prediction for all tested functionals). A second apparent discrepancy is that in the STM images, every other molecule appears to be rotated by 90°. In contrast, our DFT calculations find parallel molecules to be energetically favorable by about 80 meV, both on the surface and in the gas phase (cf. Table II in the Supporting Information [24]). The energy differences are about one order of magnitude larger than our numerical accuracy. We want to stress that the discrepancies between experiment and theory are not a shortcoming of our machine learning model, but borne out from the underlying electronic structure theory.

While the origin of this discrepancy might be ascribed to kinetic trapping or deficiencies of the underlying electronic structure method, it does not compromise the efficiency of our machine-learning approach, which truthfully reproduces the DFT PES. Moreover, it is interesting to note that all of the energetically low-lying structures that we find are variation of the ground-state structure, in particular kinks along the diagonal lines (Fig 7c) at various positions. No other defects with comparably low formation energies exists: Breaking periodicity in the high symmetry direction by introducing a line of inequivalent molecules (Fig. 7d) has an energetic cost of 120 meV per molecule in the line. A particular strength of our approach is that now, the inspection of the posterior interaction energies allow us to understand why these structures form. For the following discussion, we will focus on the "top" and the "bridge" position (geometries A and C in Fig 1), since these are the only local adsorption geometries that occur, both in the experiment and in our prediction.

For the isolated molecules, the adsorption energies $U_i$ for top and bridge are -1.81 eV and -1.51 eV, respectively. (The minus sign indicates an exothermic adsorption energy). After the training, the metal-molecule interaction is notably reduced (presumably due to depolarization). However, both adsorption geometries are affected almost equally, shifting by ca. 50 meV. Since the top site is more favorable by ca. 300 meV, if the molecules were truly non-interacting, the most favorable structure would consist of equally spaced top-sites, regardless of the preparation temperature. Since this is not the case (in experiment nor prediction), it is self-evident that intramolecular interaction plays a major role for the structure formation.

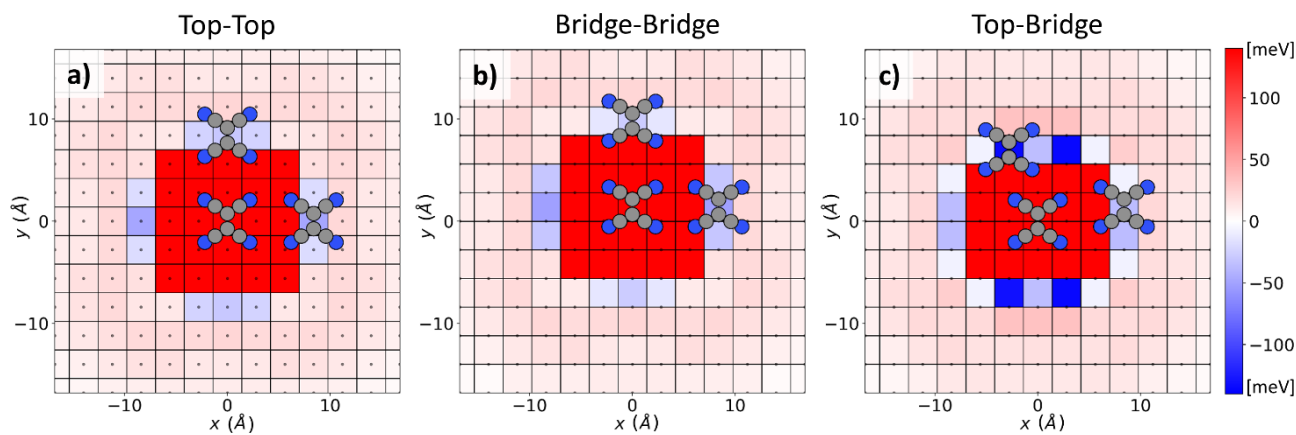

*Figure 8: Interaction energies between TCNE molecules on the Ag(100) surface. The maps shows the interaction energy $V_p$ around a fixed molecule in the center. Each box corresponds to one possible adsorption site around the central molecule, with its interaction energy for that distance color-coded. Two additional molecules located on low energy sites are depicted for better visualization of these sites.*

In Figure 8, we plot the interactions between two molecules adsorbed in top position (Fig 8a), two molecules adsorbed in bridge position (Fig 8b) and one molecule adsorbed in top and the other in bridge position (Fig 8c) as function of their separation. (Note that each box in those graphs corresponds to a separate $V_p$ parameter). In all cases, the interactions are found within a range of approximately ±150meV. First, we note that all three motifs behave qualitatively similar. Because on the surface, the TCNE molecules are negatively charged, at medium distance the interaction is mildly repulsive (red region in Fig 8), dropping towards zero at long distance. At short distances, the electrostatic repulsion is overpowered by attraction (blue region in Fig 8) due to van-der-Waals interactions. At even shorter distances, Pauli repulsion leads to rapid increase in energy, which prevents the interpenetration of two molecules.

Inspection of the top-top and bridge-bridge interaction shows that in both cases, the preferred (i.e. most attractive) interaction is found at a distance of 3 Ag lattice constants, either directly parallel or orthogonal to the C=C bond of the TCNE molecule. The minimum is similarly deep in both directions, though slightly preferred in the orthogonal direction. In contrast, the interaction between top and bridge is more anisotropic. Although again, we find minima both parallel and orthogonal to the C=C bond, they differ both in depth and position. While only a shallow minimum is found in the direction, orthogonal to the C=C bond, there are deep minima when the molecule is slightly diagonally displaced. This is, indeed, overall the most favorable interaction. As shown in Figure 8, the position of this interaction is such that the cyano-groups of the two molecules elude each other, while still allowing the molecules to come as close to each other as possible. Notably, the separation of the molecules is ca. 20% closer than at the optimal top-top interaction, allowing the overall layer to pack more densely.

The interaction maps thus allow understanding the formed structure. The preferred interaction between TCNE molecules occurs between top and bridge structures, which moreover allow the molecular layer to pack more closely than if only top geometries were present. The position of this minimum is located not directly along the crystallographic direction, but slightly shifted, which leads to the observed diagonal lines. Due to the symmetry of the lattice, there are degenerate minima both when the molecule is displaced slightly "to the left" or "to the right". It is, therefore, essentially random whether the molecules continue to grow in one line or whether they form a kink in the structure. Because the top-bridge interaction shows only twofold rotational symmetry (not fourfold, as the lattice does), the top-bridge alternation – and the kinks that come with it – are only found in one direction.

**Summary.** We have developed a machine learning model to predict the formation energies of organic monolayers. Training our model on as few as 100 DFT calculations of small periodic systems enables us to make predictions for large unit cells with DFT accuracy, enabling an extensive overview of the potential energy surface. Although our method is not necessarily cheaper than established structure search methods, it provides more relevant information (such as defect energies) for the same cost.

Applying our method to the case of TCNE on Ag(100), we find that we can accurately reproduce the results of the underlying DFT potential energy surface. We find the most stable structure to contain 6 molecules in a rectangular unit cell. Of these, only 2 are symmetry inequivalent (i.e., the unit cell could be described as a monoclinic unit cell with two molecules). The fact that we could correctly find and predict the corresponding rectangular supercell, which is three times as large, is a further indication for the performance of our method. In agreement with experiment, we predict the most stable structure to consist of diagonal lines in one direction, but straight lines in the other. Low energy defects break the periodicity along the diagonal lines, leading to the same kinked structure that is also observed experimentally. This finding underlines the importance to systematically sample low-energy structures beyond the global minimum for realistic materials.

We see applications for our method in a large variety of surface science problems, in particular for structure search, study of polymorphs and defects. Including information about transition barriers would make this model well suited for Monte Carlo studies of growth and surface dynamics due to its high accuracy at small computational cost.

**Acknowledgements.** We gratefully acknowledge Daniel Wegner for supplying experimental data and helping with their interpretation. We thank Egbert Zojer, and Matthias Rupp for fruitful discussions. Financial support by the Austrian Science Fund (FWF): P28631-N36 is gratefully acknowledged. The computational studies presented have been achieved using the Vienna Scientific Cluster (VSC) and the Argonne Leadership Computing Facility (ALCF), which is a DOE Office of Science User Facility supported under Contract DE-AC02-06CH11357.

# 1 Method Details

We used a 6 layer silver slab with a lattice constant of 3.94 Å for all surface calculations with a modified "tight" basis-set (removing Ag 5g and 4d basis functions) and an integration grid radial multiplier of 1. Our k-points were converged to a density of 24 k-points for the primitive Ag unit cell and scaled accordingly for larger unit cells. The geometry optimizations for the *local adsorption geometries* were done in a 6x6 supercell using the Broyden-Fletcher-Goldfarb-Shanno algorithm until the remaining forces were less than 0.01 eV/Å. All adsorption energies for multi-molecule configurations were obtained by single-point calculations. The machine learning was done using a custom python code using numpy, scipy and spglib. Visualizations were obtained using matplotlib and ASE.

For this study, we focus on the experimentally observed coverage for TCNE/Ag(100) of 59Å²/molecule. However, we emphasize that this is not a necessary input, since it could, in principle, also be independently determined by determining polymorphs for various coverages and finding the one with the lowest Gibb's energy per area. Furthermore, for this study we limit our search to rectangular unit cells of arbitrary size, for a simple technical reason: It allows us to systematically scale the k-point density and exploit equivalent k-points in (almost) all calculations, thus keeping the calculations numerically consistent. This facilitates the benchmark of the machine-learning model, which would be non-trivial when dealing with oblique unit cells.

## 2 D-optimal training set selection
To demonstrate the power of d-optimal training set selection we trained the model on 1000 randomly selected training sets (each containing 48 configurations) from the TCNE/vacuum test system and recorded its Root Mean Square Error (RMSE) on a validation set. Fig. 1 shows the distribution for these 1000 RMSE values compared to the RMSE of a d-optimally selected training set. The d-optimally selected set outperformed the random selection in 97% of all trials and gave a RMSE of 13 meV while the randomly selected test sets had a mean RMSE of 18 meV.

## 3 PBEsol dataset
To show that the method is transferrable between different methodologies, in addition to PBE we also obtained the potential energy surface (PES) for TCNE/Ag(100) using the PBEsol exchange-correlation functional. PBEsol yields a significantly different PES, in particular because it destabilizes the *local adsorption geometry A* ("top") relative to the other *local adsorption geometries*.

Nonetheless the machine learning model can just as well reproduce the results obtained by the PBEsol functional when trained on PBEsol calculations. Fig. 2 shows the predictions for a validation set after having trained the model on 68 configurations. Just as for the PBE dataset, also for PBEsol the prediction accuracy is high with a Root Mean Square Error of 12 meV.

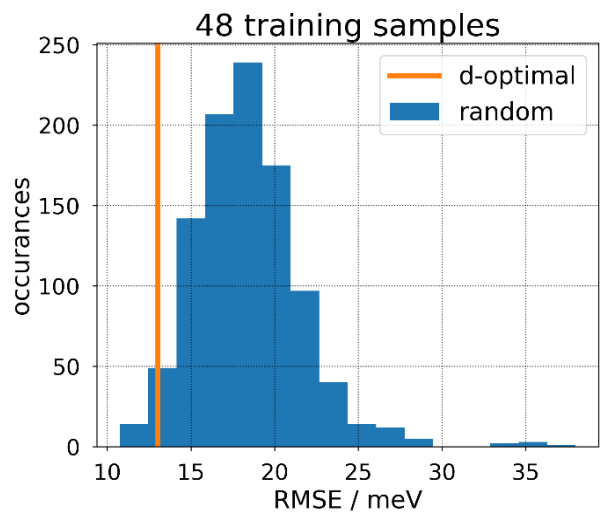

**Figure 1 |** Root Mean Square Error (RMSE) distribution for randomly selected training sets compared to the RMSE obtained by D-optimal training set selection. D-optimal selection beats random selection in 97% of trials and decreases the mean prediction error by about 30% at no additional computational cost.

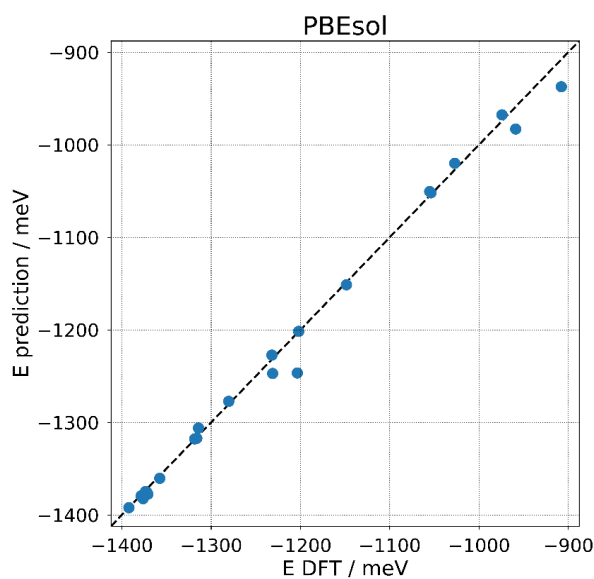

**Figure 2 |** Prediction accuracy for the PBEsol dataset. Training the model on 68 DFT calculations yields a RMSE of 12 meV on this validation set and again no significant outliers. This underlines the transferability of the machine learning model between different methodologies.

## 4 Effect of the Methodology: Top vs Bridge
To investigate the difference in adsorption energy between the top and the bridge geometry in more detail we calculated the adsorption energy for an isolated molecule with different functionals. For all functionals we calculated the adsorption energy of a molecule sitting on either a top or a bridge position using the geometries obtained from PBE. The energy differences between top and bridge are listed in Tab. 1. When including the vdW$^{surf}$ correction the top geometry is lower in energy by more than 200 meV compared to the bridge geometry, independent of the XC-functional used.

To estimate the influence of vibrational enthalpy we calculated the vibrational energy of both the *top* as well as the *bridge* adsorption geometry while keeping the positions of the substrate atoms fixed. The vibrational zero-point energy (ZPE) for the *top* geometry is 1.210 eV, the ZPE for the *bridge* position is 1.193 eV. The vibrational ZPE thus raises the adsorption energy of the *top* geometry by only 17 meV relative to the *bridge* geometry and can therefore not sufficiently destabilize the *top* adsorption geometry to account for the more frequent observation of bridge sites in experiment.

**Table 1 |** Difference in adsorption energy between the Bridge and Top adsorption geometry for different XC-functionals, including and excluding the impact of the vdW$^{surf}$ correction. and optionally TS van der Waals correction.

| functional | with vdW / meV | without vdW / meV |
|---|---|---|
| **PBE** | 316 | 138 |
| **HSE** | 393 | 211 |
| **PBEsol** | 219 | 240 |
| **revPBE** | 249 | 65 |
| **AM05** | 206 | 206 |
| **SCAN** | 224 | 224 |
| **TPSS** | 213 | 213 |

## 5 Effect of the Methodology: Parallel vs Orthogonal Molecules

To address the issue of orthogonal vs rotated molecules we calculated the energetic difference between both geometries for a polymorph with 2 molecules per unit cell with a variety of different methodologies. We always find that it is energetically favorable for the molecules to align parallel, as opposed to aligning orthogonal to each other. We observe that this energetic difference is already present when considering a TCNE dimer in the gas-phase. Furthermore, this energetic ordering is independent from the exact positioning of the molecules relative to each other. Fig. 3 shows that for all positions of the TCNE molecules relative to each other the parallel arrangement is favorable compared to the orthogonal arrangement. We also investigated this energetic difference between parallel and rotated molecules in gas-phase for different methodologies as listed in Tab. 2. None of these changes significantly altered the energetic difference: All settings resulted in the parallel orientation to be favorable by about 50-60 meV.

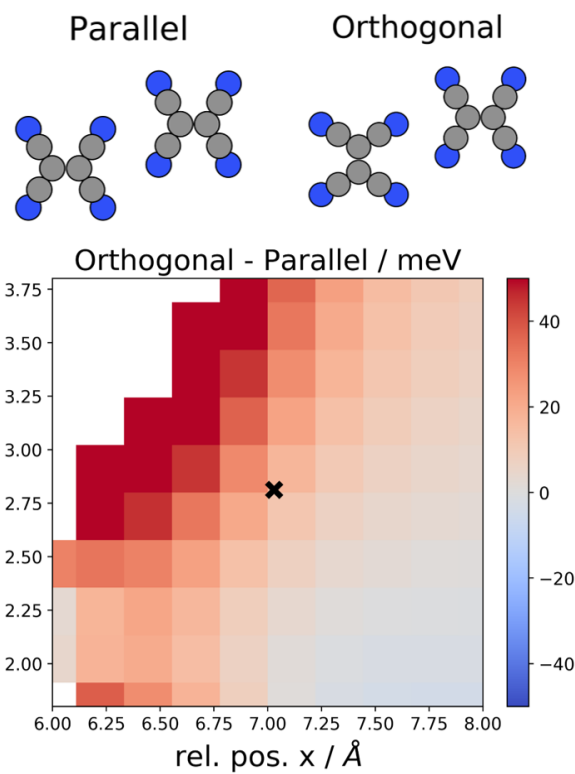

**Figure 3 |** Energetic difference between dimers of parallel and rotated molecules depending on the dimer separation. For all considered relative positions the parallelly oriented molecules are lower in energy compared to the orthogonally arranged molecules. The x marks the relative position in the periodic polymorph and the tests conducted in Tab. 2.

**Table 2 |** Energetic difference between rotated and parallelly oriented molecules for different computational settings

|  | parallel/eV | orthogonal/eV | Δ/meV |
|---|---|---|---|
| **PBE** | -24356.261 | -24356.202 | **59** |
| **PBE large basis set** | -24356.312 | -24356.253 | **59** |
| **PBE + MBD** | -24356.261 | -24356.202 | **59** |
| **PBE0** | -24354.279 | -24354.222 | **57** |
| **SCAN** | -24372.308 | -24372.254 | **54** |

.

2